\documentclass[prb,superscriptaddress,showpacs]{revtex4-1}
\usepackage{graphicx}
\usepackage{dcolumn}
\usepackage{bm}
\usepackage{amsmath}
\usepackage{amssymb}
\usepackage{ulem}
\usepackage{color}

\def\dr{\!dr\,}
\def\drp{\!dr'\,}
\def\dR{\!dR\,}
\def\dRp{\!dR'\,}
\def\du{\!du\,}
\def\dup{\!du'\,}
\def\dk{\!dk\,}
\def\dkp{\!dk'\,}
\def\dkk{\!d^2k\,}
\def\dkkp{\!d^2k'\,}
\def\dnu{\!d\nu\,}

\def\H{{\cal H}}

\def\X{{\cal X}}
\def\E{{\cal E}}
\def\C{{\cal C}}

\def\rs{{r_s^{}}}

\def\arctanh{{\tanh^{-1}}}

\def\camin{{c_{a,{\rm min}}}}
\def\cmin{{c_{\rm min}}}
\def\cmina{{c_{\rm min}^0}}
\def\umax{{u_{\rm max}}}
\def\R{\mathbb R}
\def\Vv{\hat v}

\begin{document}

\title{Quasi-two-dimensional electron gas at metallic densities}
\author{B. Bernu}
\affiliation{LPTMC, UMR 7600 of CNRS, Universit\'e P. et M. Curie, Paris, France}
\author{F. Delyon}
\affiliation{CPHT, UMR 7644 of CNRS, \'Ecole Polytechnique, Palaiseau, France}
\author{M. Holzmann}
\affiliation{LPTMC, UMR 7600 of CNRS, Universit\'e P. et M. Curie, Paris, France}
\affiliation{LPMMC, UMR 5493 of CNRS, Universit\'e J. Fourier, Grenoble, France}

\date{\today}
\begin{abstract}
We consider the three-dimensional
electron gas confined by a strictly two-dimensional  homogeneous positive charge density at $z=0$.
Within the Hartree-Fock approximation, we study the mode structure in the confined direction
in the metallic regime.
We find, that for $r_s<1.3$ ($r_s<2.5$) the unpolarized (polarized) electron gas 
starts to populate also the first excited state in the $z$-direction.
\end{abstract}
\maketitle

\section{Introduction}

The two-dimensional homogeneous electron gas (2DEG) is one of the 
most simple and thus widely used model to study 
electronic correlations in
two dimensions \cite{Rajagopal,Tanatar,Drummond}.
Experimentally,  
 two-dimensional electronic systems
have been  realized using heterostructures, e.g. semiconductor-insulator interfaces,
where layers of electrons  are tightly confined in one spatial dimension ($z$) by strong
surface electric fields, and  the discreteness of the quantized
energy levels in $z$ becomes important \cite{AndoFowlerStern}.
However, since electronic wave functions and electromagnetic
fields spatially extend in the $z$-direction, 
theoretical predictions for the 2DEG must be modified 
before a quantitative comparison is possible \cite{Saverio}.

In order to study general effects due to the interplay of electron-electron
interactions and correlations with the finite extension of the electronic density in the $z$-direction,
we introduce the model of a quasi-two-dimensional electron gas (Q2DEG). This model provides a simple and
natural extension of the 2DEG which contains essential features of
more sophisticated microscopic descriptions of heterostructures  \cite{AndoFowlerStern}.
Frequently, experiments are modeled with additional parameters to account for the finite thickness. In general, these parameters should not be considered as independent of the density due to charge neutrality.

Similar to the electron gas in two and three dimensions, we consider a jellium of electrons in
a positive charged background  insuring  total charge neutrality. Whereas the electrons are treated
fully three-dimensional, the background charges  remain strictly two-dimensional, described 
by a homogeneous charge density, $\sigma_0$, in the plane $z=0$. For vanishing total charge of the system,
the electrons are confined around the plane $z=0$.
Similar to the 2DEG, we introduce the dimensionless parameter $\rs=1/(a_B\sqrt{\pi\sigma_0})$ where $a_B=\hbar^2/(m_ee^2)$ is the Bohr radius, $m_e$ the mass and $(-e)$ the charge of the electron. At zero temperature, the system is fully described by
the value of $\rs$ which characterizes the effective two-dimensional density of the electrons.

In this paper, we study the Q2DEG in the metallic density region ($0.5 \lesssim \rs \lesssim 5$) in the Hartree-Fock approximation.
In particular we determine the spatial density distribution of the electrons in the $z$-direction, and the possible transition
between the occupation of a single confined mode to the occupation of two or more excited modes, or subbands, in $z$.
We show  that for $\rs \to 0$ the energy per particle can be written as:
\begin{eqnarray}
	E_{m}^{}(c_a,\rs)&=&\frac{K_p}{r_s^2}\sum_{a=1}^m c_a^2+\frac{\E_{m}(c_a,\rs)}{r_s^{4/3}}+\frac{\X_{m}^{}(c_a,\rs)}{\rs}+\C_m(c_a,\rs)
\end{eqnarray}
where $m$ is the number of occupied modes in the $z$-direction and $\E_{m}$ and $\X_{m}^{}$ are smooth functions of $r_s$, determined within Hartree-Fock (HF), $K_p$ is a constant for fixed spin polarization, $p$, and $c_a$ are the concentrations of electrons in each mode.
The correlation energy beyond Hartree-Fock, $\C_m$, is estimated within density-functional theory.
At fixed density (fixed $r_s$), we determine the  ground state for given concentrations, $c_a$, and,
finally,  minimize with respect to the concentrations to obtain $E_m(\camin,\rs)$.
The main goal of this paper is to determine the density where the two-mode solution (section \ref{SEC-2M})
becomes energetically favorable compared to the single mode solution.

The paper is organized as follows. Section~\ref{SEC-DEFPOT} introduces the model Hamiltonian of the Q2DEG and discusses
the technical problems related to the thermodynamic limit and the long range behavior of the Coulomb $1/r$-potential in the
potential energy. In Section~\ref{SEC-HF}, we use the HF approximation to simplify the many-body problem, and
discuss the general structure of the ground state energy in the high density limit, $r_s \to 0$. In the following sections,
Section~\ref{SEC-1M} and Section\ref{SEC-2M}, we discuss the single mode and two mode solution of the HF approximation.
For both cases, we first start discussing the Hartree-approximation, where we have found analytical solutions
for the resulting   non-linear Schr\"odinger equation. These solutions serve to obtain a first estimate for the
Hartree and exchange contribution to the 
energy, $\E_m^0$ and $\X_m^0$, respectively. We will show later, that the numerical minimization of the full HF-energy introduces only minor corrections. Finally, we briefly discuss correlation effects beyond HF
within the local density approximation using density functional theory (Section~\ref{SEC-DFT}).

\section{Quasi-two-dimensional electron gas model}
\label{SEC-DEFPOT}

Let us consider $N$ electrons interacting with a homogeneous positive charged, strictly two-dimensional plane at $z=0$ and area $S=L^2$. 
Assuming a charge-neutral system, the background surface density writes $\sigma_0=N/S$.
The $N$-body Hamiltonian is given by
\begin{eqnarray}
	\H_N=\sum_{i=1}^N -\frac{\hbar^2}{2m_e}\Delta_i +V_N
\end{eqnarray}
where  $V_N$ is the total potential energy of all charges.

It is well known that the Coulomb potential poses difficulties 
in the definition of the potential energy in the thermodynamic limit due to the non integrability at infinity. The local singularity of the Coulomb potential near the origin is a classical problem of self-adjointness and here we 
only focus on the definition of the potential with periodic boundary conditions.

Let $\Lambda$ denotes the two-dimensional lattice in $\R^3$ generated by the vectors $(L,0,0)$ and $(0,L,0)$. For a regular integrable interaction $v$, we formally define the total periodic potential as:
\begin{align}
	V_N&=V_{ee}+V_{eb}+ V_{bb} \\
\label{Vee}
	V_{ee}& = \frac 1 2\sum_{i\neq j,\tau\in\Lambda}v(R_i-R_j+\tau)+ \frac 1 2\sum_i\sum_{\tau\in\Lambda,\tau\neq 0}v(\tau)\\
	V_{eb} &= -\frac N S \sum_{i}\int_{\R^2} \dr v(R_i-r)\\
	V_{bb} &=  \frac{N^2}{2S^2} \int_{S\times \R^2}\dr\,\drp v(r-r')
\end{align}
where the index $e$ holds for the electrons and $b$ holds for the positive background. 
The last term in Eq.\ref{Vee} is the interaction of an electron with all its periodized images.
As soon as the interaction $v$ is regular and integrable, we can rewrite the potential energy as:

\begin{align}
\label{EQ-VP}
	V_N=&\frac 1 2\sum_{i\neq j}\left(\Vv(R_i-R_j)+\frac1S v_1(z_i-z_j)\right)-\sigma_0\sum_i v_1(z_i)+\frac N 2C_v
\end{align}
with
\begin{align}
\Vv(R)
\label{EQ-Vv}
&=\sum_{\tau\in\Lambda}\left(v(R+\tau)-\frac 1 S \int_{S}\dr v(R+\tau+r)\right)
\\
	v_1(z)&=\int_{\R^2} \dr v((r,z))-v((r,0))
\end{align}
and $C_v$ is the Madelung energy of electrons on the lattice $\Lambda$ in a homogeneous background
\begin{align}
	C_v&=\sum_{\tau\in\Lambda,\tau\neq 0}\left(
			v(\tau)-\frac 1 S \int_{S} \dr v(r+\tau)
		\right)-\frac 1 S \int_{S} \dr v(r)
\end{align}
Let us notice that the Fourier transform $\tilde \Vv(K)$ of $\Vv$ is directly related to the Fourier transform, $\tilde v(K)$,  of $v$.
As can be directly verified, we have  $\tilde \Vv(K)=\tilde v(K)$, except that $\tilde \Vv(K)=0$ for $k_x=k_y=0$.
With this new definition (Eq.\ref{EQ-VP}), we only need that $\Vv$, $v_1$ and $C_v$ are well defined.
That is :
\begin{align}
	\sum_{\tau\in\Lambda}\left|v(R+\tau)-\frac 1 S \int_{S}\dr v(R+\tau+r)\right|<+\infty\\
	\int_{\R^2}\dr |v((r,z))-v((r,0))|<+\infty
\end{align}
These conditions are fulfilled by the Coulomb potential $v_C(R)=e^2/R$, except at $R=0$ as mentioned above.
Furthermore, we have $v_1(z)=-2\pi e^2 |z|$ and $\tilde \Vv(K)=4\pi e^2/(k_x^2+k_y^2+k_z^2)$ for $k_x\neq 0$ or $k_y\neq 0$ and 0 otherwise,  and the periodic potential energy,  Eq.~(\ref{EQ-VP}),  finally writes
\begin{eqnarray}
\label{EQ-defVN}
	V_N&=&\frac{e^2}2\sum_{i\ne j}\left[v^{q2D}(R_{ij})
		-\frac{2\pi\left|z_i-z_j\right|}S \right]
		+2\pi e^2\sigma_0\sum_i \left|z_i\right|+\frac N2 C_v\\
	v^{q2D}(R)&=&\frac1S\sum_{k\ne0}\int\frac{dk_z}{2\pi} \frac{4\pi}{k^2+k_z^2}e^{iK\cdot R}
\end{eqnarray}
where $K=(k,k_z)$ and $e^{ik\cdot \tau}=1$ for $\tau\in\Lambda$.

\section{Hartree-Fock Approximation}
\label{SEC-HF}

Within the HF approximation we minimize the  ground state energy per particle, $E$,
with respect to variations of the many-body wave-function, $\Psi_N=\det \left|\left\{\Psi_{i\uparrow}\right\}\right|\det \left|\left\{\Psi_{i\downarrow}\right\}\right|$, in the subspace of single
Slater determinants
\begin{equation}
	E=\frac1N\frac{\left<\Psi_N\left|\H_N\right|\Psi_N\right>}{\left<\Psi_N|\Psi_N\right>}
\end{equation}
 In the following  we assume that $\left\{\Psi_{i\sigma}\right\}$ ($\sigma=\uparrow,\downarrow$) are normalized, orthogonal single particle wave-functions, 
 and
 we obtain for the total energy per particle:
\begin{eqnarray}
	E&=&
	-\frac1N\sum_{i\sigma} \int_{S\times \R} \dR \Psi_{i\sigma}^*(R)\frac{\hbar^2 }{2 m_e}   \Delta \Psi_{i\sigma}(R)
	\nonumber\\
	&&+\frac1N\frac{e^2}2  \int_{S\times \R} \dR\, \dRp n_e(R)v^{q2D}(R-R') n_e(R')
	\nonumber\\
	&&-\frac1N\frac{e^2}2  \sum_{i,j,\sigma}\int_{S\times \R} \dR\, \dRp \Psi_{i\sigma}^*(R)\Psi_{j\sigma}^{}(R)v^{q2D}(R-R')\Psi_{i\sigma}^{}(R')\Psi_{j\sigma}^{*}(R')
	\nonumber
	\\
	&&+\frac1N\int_{S\times \R} \dR n_e(R)2\pi e^2\sigma_0|z|
		-\frac{\pi e^2} {NS}\int_{S\times \R} \dR\, \dRp n_e(R)|z-z'| n_e(R')
		\nonumber\\
		&&+\frac{\pi e^2} {NS}\sum_{i,j,\sigma}\int_{S\times \R} \dR\, \dRp \Psi_{i\sigma}^*(R)\Psi_{j\sigma}^{}(R)|z-z'|\Psi_{i\sigma}^{}(R')\Psi_{j\sigma}^{*}(R') + \frac{C_v}{2}
		\label{EQ-ETHFmulti}
\end{eqnarray}
where we have defined the total electronic density by
\begin{equation}
\label{EQ-rhomulti}
	n_e(R)=\sum_{i\sigma} \left|\Psi_{i\sigma}(R)\right|^2
\end{equation}

In this paper we are interested in a quasi-two-dimensional regime where we expect that the electrons populate 
a finite number, $m$, of discrete modes in the $z$-direction, whereas the density of states is continuous in the
plane at constant $z$, in the thermodynamic limit.
Each single-body wave function $\Psi_{i\sigma}$ is then taken as a product of a plane wave $\phi_k$ in the plane $z=0$ and a wave function $\psi_a$ in the $z$-direction where $a$ labels the mode.
Let $N_a$ be the number of electrons in the mode $a$ and $c_a^{}=N_a^{}/N$, $N=\sum_a N_a^{}$.
Accounting for the spin polarization $p$ of the electrons, we have $N_a^{}=N_{a\uparrow}^{}+N_{a\downarrow}^{}$ and $c_a^{}=c_{a\uparrow}^{}+c_{a\downarrow}^{}$.
In the following, we restrict the discussion to 
$i)$ the fully polarized gas ($p=1$) where $c_a=c_{a\uparrow}$ and $c_{a\downarrow}=0$,
and $ii)$ the unpolarized electron gas ($p=0$) with $c_{a\uparrow}=c_{a\downarrow}=c_a/2$ (unpolarized in each mode $a$).
We further assume that the wave functions do not depend on spin: $\psi_{a\sigma}\equiv \psi_a$.

In each mode $a,\sigma$, all transverse plane waves  are occupied up to $k_{F_a\sigma}^{}=\sqrt{c_{a\sigma}^{}}k_F^{}$ with $k_F^{}a_B^{}=2/r_s^{}$, and,
in the thermodynamic limit,  all summations over transverse states are replaced by integrals inside the Fermi surfaces
\begin{eqnarray}
\sum_{i\sigma}\equiv\sum_{a\sigma}\sum_{|k|<k_{F_{a\sigma}}}
	&\longrightarrow&\sum_{a\sigma}\frac{N}{\pi k_{F}^2}\int_{|k|<k_{F_{a\sigma}}}\!\!\!\!\!\!\!\!\!\!\!\!d^2k.
\end{eqnarray}
Further, for $N\to \infty$,  the last line of Eq.~(\ref{EQ-ETHFmulti}) vanishes.

It is instructive to regroup the different contributions to the total energy (in Hartree) as follows
\begin{equation}
\label{EQ-defet}
	E[c_a,\psi_a,\rs]=
	\frac{K_p \sum_a c_{a}^2 }{r_s^2} +\frac{\E[c_a,\psi_a]}{r_s^{4/3}} +\frac{\X[c_a,\psi_a,\rs]}{r_s^{}} 
\end{equation}
where the first term is the in-plane, strictly two-dimensional,  kinetic energy with $K_0=1/2$ for the unpolarized and $K_1=1$ for the polarized electron gas.
In order to separate the explicit $r_s$-dependency in the following two terms, we introduce  $u=r_s^{1/3} k_Fz$ together with the normalization  $\int_\R\du|\psi_{a\sigma}^{}(u)|^2=1$ of the confined modes.
All contributions independent of the in-plane modes are contained in
 $\E$
\begin{equation}
\label{EQ-defeHx}
	\E[c_a,\psi_a]=-2\sum_{a} c_{a}^{}\int_\R\du \psi_{a}^{}(u)\psi_{a}^{''}(u)
		+\int_\R\du \rho(u)\frac{|u|}2+\int_\R\du \rho(u)v_\rho^{}(u)
\end{equation}
where the electrostatic potential, $v_\rho$, is determined by the one-dimensional Poisson equation
\begin{equation}
\label{EQ-defvs}
	v_\rho''(u)=\delta(u)-\rho(u)
\end{equation}
from the total electronic density distribution $\rho(u)=\sum_{a\sigma} c_{a\sigma} |\psi_{a\sigma}|^2=\sum_{a} c_{a} |\psi_{a}|^2$ and the positive background charges at $z=0$. Using $v_\rho(\infty)=v_\rho'(\infty)=0$, we have
\begin{equation}
\label{EQ-defvfromrho}
	v_\rho(u)=\frac{|u|}2-\frac12\int_\R\dup\rho(u')|u-u'|
\end{equation}
The exchange term, $\X$, explicitly mixes transverse and confined states,
\begin{eqnarray}
\label{EQ-defeX}
	\X[c_a,\psi_a,\rs]&=&-\sum_{a,b}\frac{r_s^{1/3}}{4\pi}\int_\R\dnu |\tilde\rho_{ab}^{}(\nu)|^2 
		\tilde Y(c_{a}^{},c_{b}^{},r_s^{1/3} G_p \nu )
\end{eqnarray}
where $\tilde\rho_{ab}^{}(\nu)=\int_\R\du \rho_{ab}(u)\exp(-i\nu u)$, $\rho_{ab}(u)=\psi_a(u)\psi_b(u)$,
$G_0=\sqrt2$ (unpolarized) and $G_1=1$ (polarized).
The exchange function $\tilde Y$  (see Appendix \ref{APP-Y}) is given by
\begin{eqnarray}
\nonumber
	\tilde Y(c_a,c_b,\nu)&=&
			\frac2{\pi^2}\int_{|k|^2<c_a}  \!\!\!\!\!\!\!\!\! \dkk
				\int_{|k'|^2<c_b}  \!\!\!\!\!\!\!\!\! \dkkp
				\frac{1}{|k-k'|^2+\nu^2}
\end{eqnarray}
and introduces a smooth variation in $\X$ as a function of $\rs$.

The Hartree-Fock ground state is determined by minimizing the total energy, Eq.~(\ref{EQ-defet}),
with respect to $p$, $c_a$, and $\psi_a$, at fixed density, $r_s$. We simplify this rather complex 
optimization problem, by considering only the completely polarized or unpolarized electron gas.
For  fixed concentrations, $c_a$,
the minimum of $E$ with respect to $\psi_a$ is independent from the in-plane kinetic energy.
From the formal  variation of the energy with respect to $\psi_a$ we obtain
\begin{eqnarray}
\label{EQ-defHd1}
	\frac{	dE}{d \psi_a}
	&=& \frac{4}{r_s^{4/3}} \H_0c_a \psi_a + \frac{4}{r_s}  \sum_{b} V_{a,b}^{\rm exc}(u)\psi_b 
		 \\
\label{EQ-defHd}
	\H_0&=&-\partial_u^2+v_\rho^{}(u)
			\\		
	V_{a,b}^{\rm exc}(u)	&=&-\frac{r_s^{1/3}}{4\pi } \int_\R \dnu \tilde\rho_{ab}^{}(\nu) \tilde Y(c_{a}^{},c_{b}^{},r_s^{1/3} G_p\nu)e^{i\nu u}
\end{eqnarray}

In the limit of small $r_s$, the exchange energy is negligible, and the wave-functions, 
$\psi_a \equiv \psi_a^0$ are entirely determined by minimizing $\E$, or,
equivalently by the Hartree-equation
\begin{equation}
\label{EQ-deflambda}
	\H_0 \psi_a^0 = -\lambda_a \psi_a^0
\end{equation}
which leads to
\begin{eqnarray}
\label{EQ-defEm0}
         E_{m}^0 &=&
       		\frac{K_p}{r_s^2} \sum_a c_a^2 + \frac{\E_m^0}{r_s^{4/3}}+\frac{\X_{m}^0}{r_s} 
\end{eqnarray}
where  $\X_{m}^0 \equiv \X[c_a,\psi_a^0,r_s]$ and
\begin{eqnarray}
  \label{EQ-defeH0}
	\E^0_m&\equiv& 
		\E[c_a,\psi_a^0] =-2\sum_{a} c_a^{}\lambda_a -v_\rho(0)-\int_\R\du \rho(u)v_\rho^{}(u)	
\end{eqnarray}
is independent of $r_s$.
This provides us with a semi-analytical approximation for the total energy, $E_{m}^0$, which appears to be
very close to the full minimization of the energy including exchange, $E_m$, for the densities considered.
Whereas the in-plane kinetic energy term does not influence the shape of the distribution in $z$, it favors multi-mode
occupation in the high density limit, $r_s \to 0$.

\section{Single mode solution}
\label{SEC-1M}

For a single mode, we minimize Eq.~(\ref{EQ-defet}) with
 $\rho(u)=\psi_0^2(u)$ ($c_0=1$), so that  $\psi_0$ satisfies the non-linear
 Schr\"odinger equation, Eq.~(\ref{EQ-defHd1}),
\begin{eqnarray}
\label{123}
	( \H_0+{r_s^{1/3}} V_{00}^{exc}) \psi_0= -\lambda_0 \psi_0
\end{eqnarray}
and we obtain
\begin{eqnarray}
\label{EQ-defet1M}
	E_1&=&  \frac{K_p}{r_s^2}  + \frac{\E_1}{r_s^{4/3}}+\frac{\X_{1}}{r_s} 
\end{eqnarray}
where $\E_1$ and $\X_1$ are the values of the Hartree and exchange term using the optimal $\psi_0$.

For the strictly two-dimensional electron gas, we have $\rho^{(2D)}(u)=\delta(u)$. Neglecting the zero-point
energy of the confinement, $\E_1$, in this limit,  and  using $\int_\R \dnu \tilde Y(1,1,\nu)=32/3$, 
we recover $E^{(2D)}(p=0)=1/(2r_s^{2})-8/(3\pi r_s^{}\sqrt2)$ for the unpolarized  and $E^{(2D)}(p=1)=1/r_s^{2}-8/(3\pi r_s^{})$ for the polarized electron gas.

\subsection{Hartree solution without exchange, $\E_1^0$}
\label{SEC-1MA}
We determine the one-mode solution of the Hartree-equation, Eq.~(\ref{123}) with $V_{00}^{exc} \equiv 0$,
which determines the density distribution of the mode with $\lambda_0>0$  in the high density region, $r_s \to 0$.
Assuming $\psi_0(u)$ to be an even function of $u$, we restrict the discussion to $u>0$ in the following.
From the leading order behavior $X_0$ of the solution at large $u$ where $v_\rho(u)$ vanishes, we use a series in $X_0$ as 
ansatz for $\psi(u)$,  
\begin{eqnarray}
\label{EQ-defX0}
	X_0&=&\sqrt {f_0} e^{-\alpha u}\\
\label{EQ-defpsi0}
	\psi_0(u)&=&\alpha^2\sum_{k\ge0} (-1)^ka_k X_0^{2k+1}
\end{eqnarray}
where $\alpha=\sqrt{\lambda_0}$, $f_0$, and $a_k$  are to be determined ($a_0=1$).
The density is then given by
\begin{eqnarray}
\label{EQ-rhos}
	\rho(u)&=&\alpha^4\sum_{k\ge0} (-1)^k\rho_k X_0^{2k+2}\quad{\rm with}\quad
	\rho_k=\sum_{j=0}^k a_j a_{k-j}
\end{eqnarray}
and the potential is obtained by integrating twice $v_\rho''(u)=-\rho(u)$ for $u>0$ with the conditions $v_\rho(\infty)=0$ and $v_\rho'(\infty)=0$:
\begin{eqnarray}
\label{EQ-vs}
	v_\rho(u)&=&-\alpha^2\sum_{k\ge0} (-1)^k v_k X_0^{2k+2}\quad{\rm with}\quad
	v_k=\frac{\rho_k}{4(k+1)^2}
\end{eqnarray}
and
\begin{eqnarray}
	\psi_0''(u)&=&\alpha^4\sum_{k} (-1)^ka_k(2k+1)^2 X_0^{2k+1}\\
	v_\rho(u)\psi_0(u)&=&-\alpha^4\sum_{k\ge0} (-1)^k w_k X_0^{2k+3}\quad{\rm with}\quad
	w_k=\sum_{j=0}^k v_j a_{k-j}
\end{eqnarray}
Thus, imposing $\psi_0''-(v_\rho+\alpha^2)\psi_0=0$ leads to the equation:
\begin{eqnarray}
\label{EQ-defai}
	0&=&\alpha^4X_0\left[
		\sum_{k\ge0} (-1)^k a_k [( 2k+1)^2 -1 ] X_0^{2k} +\sum_{k\ge0} (-1)^k w_k X_0^{2k+2}
	\right]
\end{eqnarray}
From the definition $a_0=1$, we get $\rho_0=1$,  $v_0=\frac14$ and $w_0=\frac14$.
The other terms are obtain by recurrence:
\begin{eqnarray}
\label{EQ-recuai}
	a_k &=& \frac{w_{k-1}}{4k(k+1)}
\end{eqnarray}
With these definitions all coefficients $a_k$, $\rho_k$, $v_k$ and $w_k$ are positive.
The two parameters $\alpha$ and $f_0$ are determined by imposing  $\psi_0'(0)=0$ and the normalization:
\begin{eqnarray}
\label{EQ-psi0constr}
	0&=&\sum_{k\ge0} (-1)^k(2k+1)a_k f_0^{k}\propto\psi_0'(0)\\
	\frac12&=&\alpha^3\sum_{k\ge0}\frac{(-1)^k\rho_kf_0^{k+1}}{2(k+1)} = \int_0^\infty \du \psi_0^2(u)
\end{eqnarray}
The numerical results are given in table \ref{TAB-1M}, together with  
the values of the different contributions to the Hartree-energy.
In particular, from Eqs.~(\ref{EQ-defHd}),(\ref{EQ-deflambda}), we have $\lambda_0=\alpha^2=-(\left<-\Delta\right>+\left<v_\rho\right>)$ 
and
from Eq.~(\ref{EQ-defeH0}), we have
$\E_1^{0}=-2\lambda_0-v_\rho(0)-\left<v_\rho\right>$ with
 $\left<v_\rho\right>=\int_\R\du\rho(u)v_\rho(u)=-\alpha^5\sum_{k\ge0}(-1)^k\tau_k f_0^{k+2}/(k+2)$, 
$ \tau_k=\sum_{j=0}^k \rho_j v_{k-j}$, and the kinetic energy writes
$\left<-\Delta\right>=\left<\psi_0\left|-\partial_u^2\right|\psi_0\right>=\alpha^5\sum_{k\ge0} \tau_k'(-f_0)^{k+1}/(k+1)$
with $\tau_k'=\sum_{j=0}^k (2j+1)^2a_j a_{k-j}$. 
Notice, that $\E_1^0$ is independent of the polarization.

\begin{table}[htdp]
\caption{Parameters and various quantities of the single mode solution of the Hartree-equation without exchange term.
}
\begin{center}
\begin{tabular}{c|c}
	$f_0$&15.5610024546998\\
	$\alpha$&0.465180466326271\\
	\hline
	$\lambda_0$& 0.216392866251527\\
	$v(0)$& -0.674164469749883\\
          $\psi_0(0)$&0.522553284700250\\
          $\left<v\right>$& -0.307947186951202\\
          $\left<-\Delta\right>$& 0.0915543206996701\\
          $\E_1^{0}$&0.549325924198031
\end{tabular}
\end{center}
\label{TAB-1M}
\end{table}

\begin{figure}
\begin{center}
\includegraphics[width=0.5\textwidth]{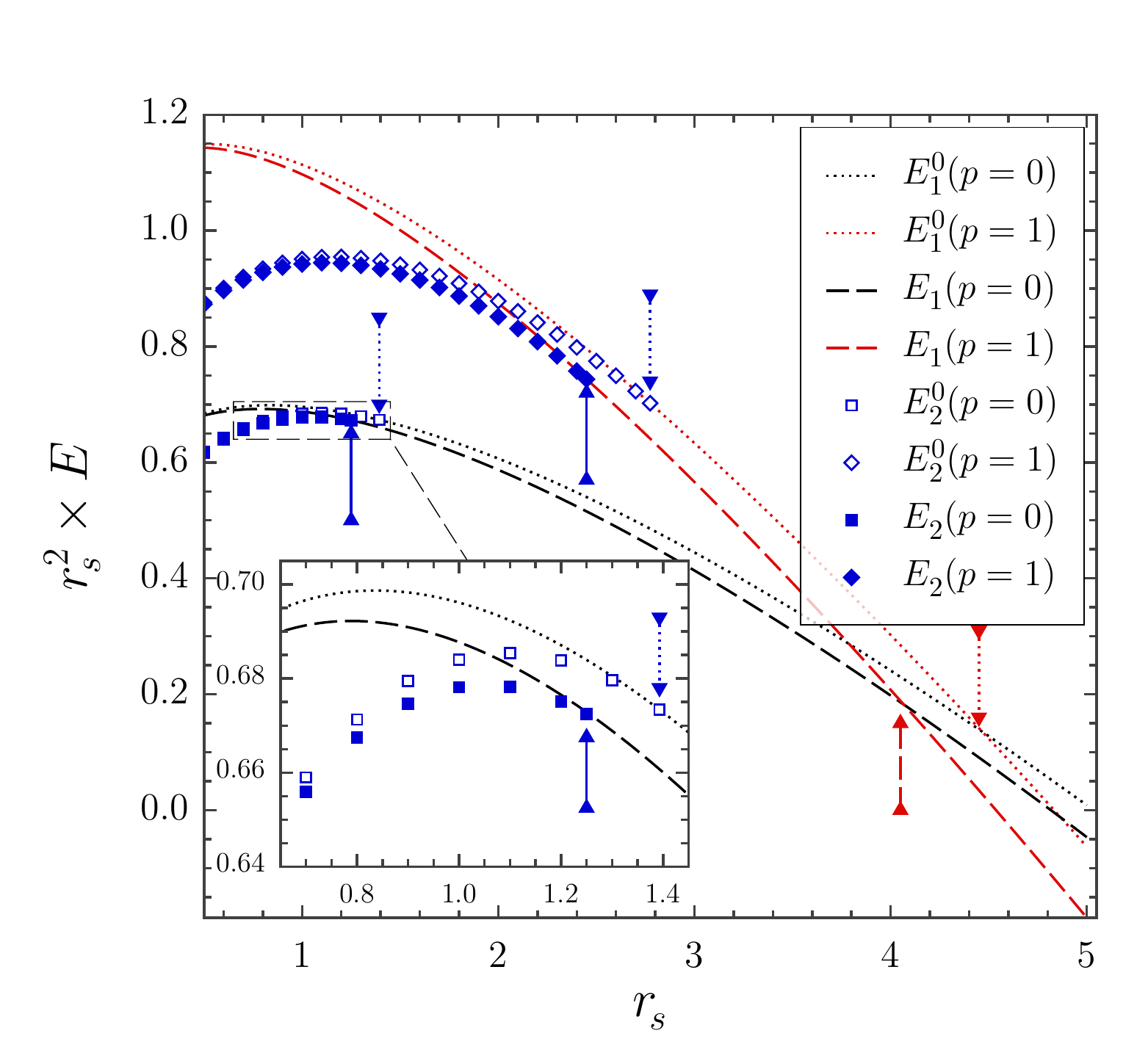}
\caption{
Comparison of the energies of the Q2DEG in the different phases within HF:
black for the single mode unpolarized ($p=0$) electron gas,
red for the single mode polarized ($p=1$) electron gas and 
blue for two occupied excited modes in $z$. 
For each phase we compare the HF energy using the Hartree density profile in $z$ with
the full HF minimization:
dashed (resp. dotted) lines stand for $E_1$ (resp. $E_1^0$) of the single mode solution  from Eq.\ref{EQ-E1M-2} (resp. Eq.\ref{EQ-defEm0}),
filled (resp. open) symbols stand for the energies including two occupied modes, $E_2$ (resp. $E_2^0$) from Eq. \ref{EQ-e2M2} (resp. Eq. \ref{EQ-e2M1}),
with squares (resp. diamonds) for the unpolarized (resp. polarized) gas.
The red arrows indicate the transition between the unpolarized gas and the polarized gas at $r_s\simeq4.45$ 
in the approximation using the Hartree density profile; minimization of the full HF energies shifts the transition to
slightly higher density,  $r_s \simeq 4.05$.
Blue arrows indicate the transitions from  the single mode system to two occupied excited modes increasing the density.
The inset shows the transition region of the unpolarized gas.
}
\label{FIG-energies}
\end{center}
\end{figure}

\subsection{One-Mode exchange-energy in the Hartree-approximation, $\X_1^{0}$}
\label{SEC-1MEXCH}
From the Fourier transform, $\tilde\rho(\nu)$,
of the ground state density, $\rho_0=\psi_0^2$,  obtained from the Hartree equation, we can
estimate  the exchange contribution, Eq.~(\ref{EQ-defeX}), to the total energy.
Since we have $\tilde Y>0$ and $0<\tilde\rho(\nu)<1$, the exchange energy of the
quasi-two-dimensional gas is greater than its strictly two-dimensional value obtained with $\tilde\rho^{(2D)}(\nu)=1$.

The main contribution of the exchange-integral comes from the logarithmic singularity of the integrand at $\nu=0$;
details on the numerical  evaluation  are given in Appendix \ref{APP-exch1M} and the results for the total energy are shown in Fig.\ref{FIG-energies}.
For densities corresponding to $0.5 \le r_s\le 5$, the exchange integral, $\X_1^0$,
is well approximated by  $\X_{1}^{0}(p=1,\rs)\approx -0.4356-0.06127\,\ln(\rs)$ for the polarized gas. 

Within the Hartree-approximation,  $\tilde\rho(\nu)$ is independent of $r_s$ and polarization, $p$,
so that a simple relation between $\X_1^0$ of the polarized and unpolarized electron gas
at different $r_s$ can be established
 \begin{eqnarray}
\label{EQ-relXUP}
\X_{1}^{0}(p=0,\rs)=\frac{\X_1^{0}(p=1,2\sqrt2r_s^{})}{\sqrt2}.
\end{eqnarray}
Using our approximate expression for $\X_{1}^{0}(p=1,\rs)$ together with
Eq.~(\ref{EQ-relXUP}) in Eq.~(\ref{EQ-defet1M}), we can
estimate, that for $\rs \gtrsim  4.56$ the polarized phase is energetically favorable
compared to the unpolarized phase.

\begin{figure}
\begin{center}
\includegraphics[width=0.5\textwidth]{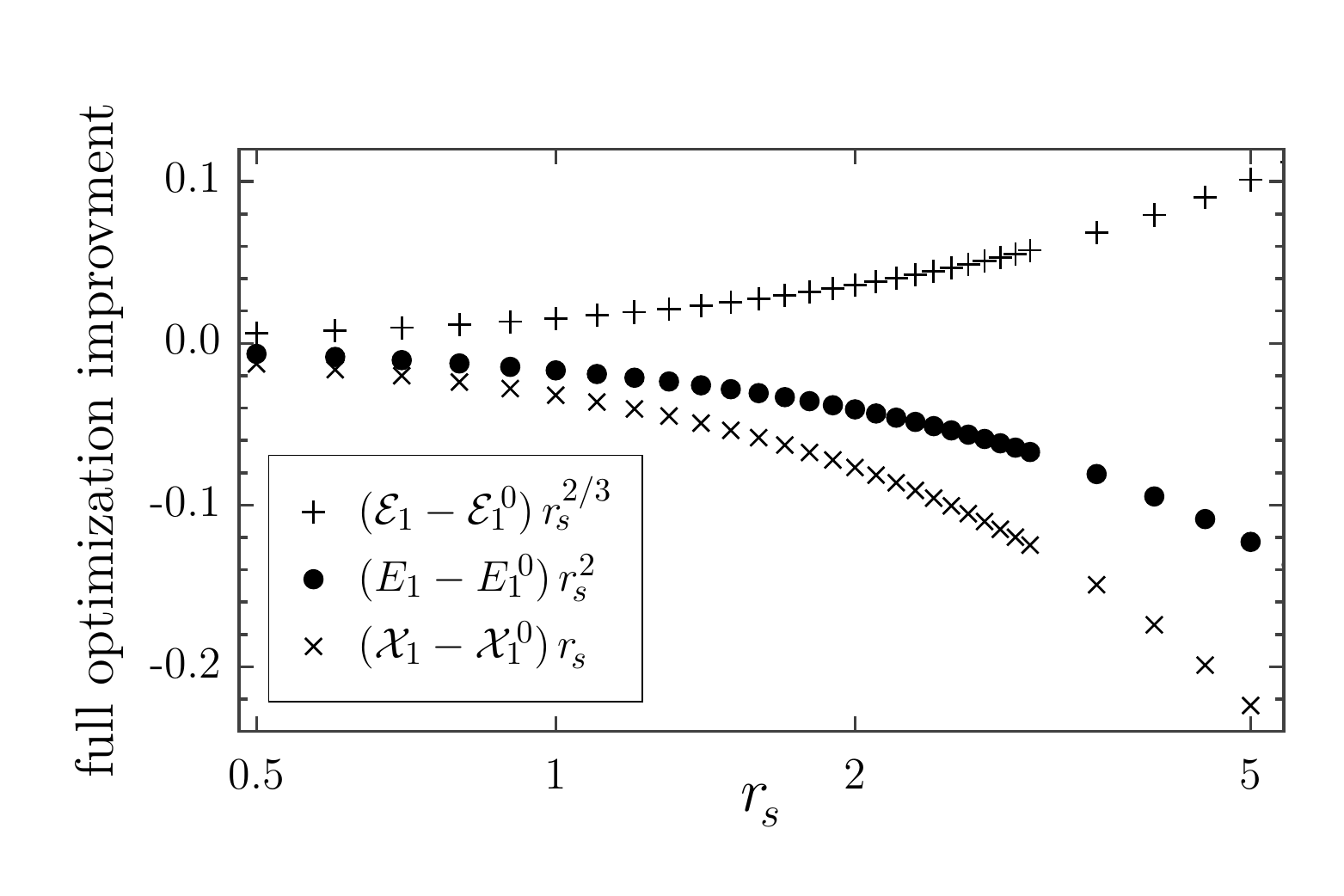}
\caption{Importance of the the full minimization of all the different components of the HF energy, from Eqs.\ \ref{EQ-defeH0} and \ref{EQ-E1M-2} for the polarized single mode gas. 
Shown are the total energy in Hartree times $r_s^2$,
together with the Hartree and exchange contributions in the same units.
The gain in exchange energy is roughly twice the increase of the Hartree energy. 
}
\label{FIG-comp1Menergies}
\end{center}
\end{figure}

\subsection{Full Minimization}
\label{SEC-1MFULL}
The full minimization 
assuming a single mode is done numerically (see Appendix \ref{APP-NUM} for the numerical details), and we have:
\begin{eqnarray}
\label{EQ-E1M-2}
	E_{1}=\frac{K_p}{r_s^2} + \frac{\E_{1}(r_s)}{r_s^{4/3}} +\frac{\X_{1}(r_s)}{r_s}
\end{eqnarray}
where $\E_1$ and $\X_1$ depend on $r_s$ and the polarization, $p$.
Figure~\ref{FIG-comp1Menergies} illustrates the small improvements due to the full minimization
compared to the Hartree-approximation, $E_1^0$.

\section{Two-mode solution}
\label{SEC-2M}
In this section we look  for the ground state energy with two modes, 
where the density is given by 
\begin{eqnarray}
	\rho(u)&=&(1-c)\psi_0^2(u)+c\,\psi_1^2(u)
\end{eqnarray}
and $c\equiv c_1$  is the concentration of the excited mode, $\psi_1$.
Analogous to the discussion of the single-mode solution,
we first minimize the Hartree-energy for given $c$ with respect to $\psi_0$ and $\psi_1$ 
to obtain $\E_2^0$.
Then, we evaluated the exchange term within this solution, $\X_2^0$, and, finally, 
we minimize the full Hartree-Fock energy including the exchange.

\subsection{Two mode Hartree solution without exchange, $\E_2^0$}
\label{SEC-2MH}

Generalizing the single mode solution of the previous section, we express the wave functions as series of exponentials.
Assuming $\psi_0(u)$ (resp. $\psi_1(u)$) to be an even (resp. odd) function of $u$,  we restrict
 $\psi_0$ and $\psi_1$ to non-negative arguments in the following 
\begin{eqnarray}
\label{EQ-def2X0}
	X_0&=&\sqrt {f_0} e^{-\alpha u}\\
\label{EQ-def2X1}
	X_1&=&\sqrt {f_1} e^{-s\alpha u}\\
\label{EQ-def2psi0}
	\psi_0(u)&=&\frac{\alpha^2}{\sqrt{1-c}}\sum_{k,k'\ge0} a_{k,k'}^{} X_0^{2k+1} X_1^{2k'}\\
\label{EQ-def2psi1}
	\psi_1(u)&=&\frac{s^2\alpha^2}{\sqrt{c}}\sum_{k,k'\ge0} b_{k,k'}^{} X_0^{2k} X_1^{2k'+1}
\end{eqnarray}
where $\alpha^2=\lambda_0$, $s^2\alpha^2=\lambda_1$ and $a_{0,0}^{}=b_{0,0}^{}=1$.
As shown in Appendix \ref{APP-2Mrec},
the coefficients $a_{k,k'}^{}$ and $b_{k,k'}^{}$  are  functions of $s$ only and can be determined by recurrence relations.
Imposing the boundary conditions at $u=0$: $\psi'_0(0)=0$ and $\psi_1(0)=0$ provide two equations independent of $c$ and $\alpha$:
\begin{eqnarray}
	0&=&\sum_{k,k'\ge0} a_{k,k'}^{}(2k+1+2k's)f_0^{k}f_1^{k'}\\
	0&=&\sum_{k,k'\ge0} b_{k,k'}^{}f_0^{k}f_1^{k'}
\end{eqnarray}
In practice the series are restricted to $k+k'\le n$.
At large enough $n$, for fixed $s$, this system of the variables $\{f_0,f_1\}$ has only one converging solution for positive $f_0$ and $f_1$.
The convergence with $n$ depends on $s$.
Relative convergence of one percent is reached at order $n \simeq 40$. 
This slow convergence is due to the difficulty to fulfill the conditions at $u=0$ as we get close to the radius of convergence of these series.
Machine precision is obtained using $n\simeq120$.

Then the normalizations of $\psi_0$ and $\psi_1$ lead to  two simple equations determining $\alpha$ and $c$:
\begin{eqnarray}
\label{EQ-defnorm0}
	\frac12&=&\frac{\alpha^3}{1-c}\sum_{k,k'\ge0}  \frac{\rho_{k,k'}^{(0)}f_0^{k+1}f_1^{k}}{2(k+1+k's)}\\
\label{EQ-defnorm1}
	\frac12&=&\frac{s^4\alpha^3}{c}\sum_{k,k'\ge0}  \frac{\rho_{k,k'}^{(1)}f_0^{k}f_1^{k'+1}}{2(k+(k'+1)s)}
\end{eqnarray}
where $\rho_{k,k'}^{(a)}$ is defined in Eq.\ref{EQ-defrhoa}.

\begin{figure}
\begin{center}
\includegraphics[width=0.49\textwidth]{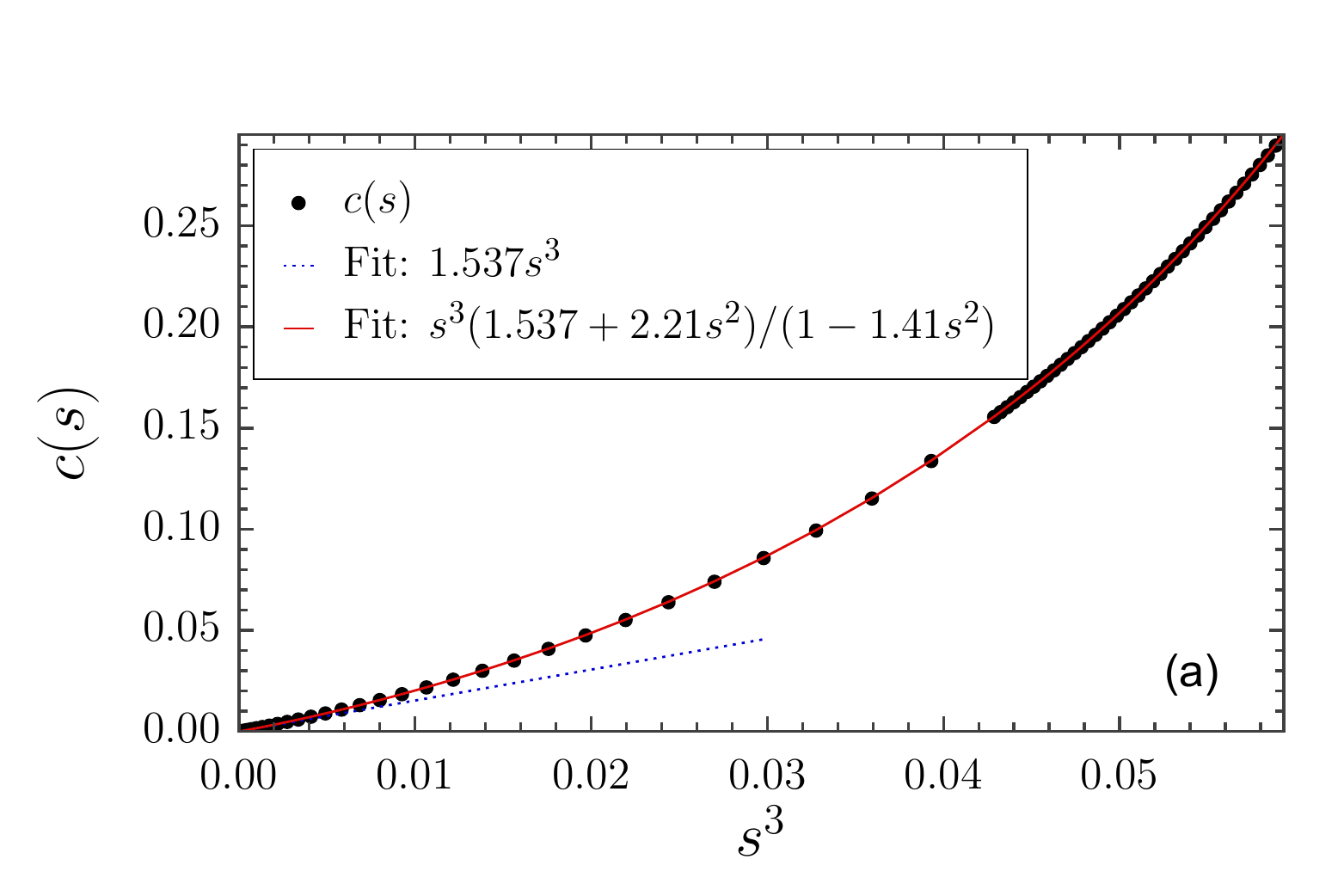}
\includegraphics[width=0.49\textwidth]{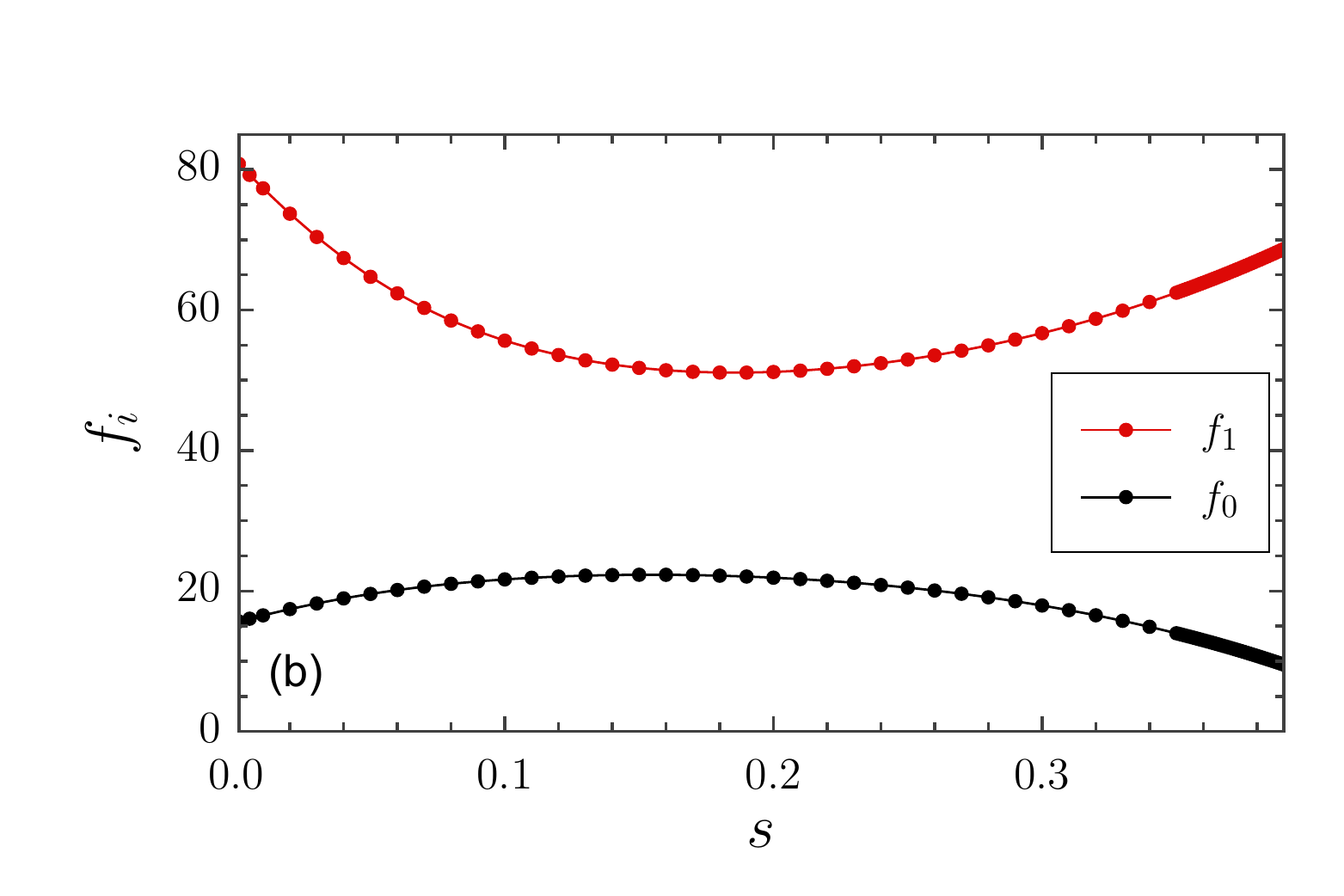}
\caption{
Graphical representation of the parameters for the two-mode model, see Eqs.\ \ref{EQ-def2X0}-\ref{EQ-def2X1}, as functions of $s=\sqrt{\lambda_1/\lambda_0}$ with $\lambda_i$ the eigenvalues of $\H_0$, Eqs.\ \ref{EQ-deflambda} . (a): concentration $c\equiv c_1$ in the first excited mode versus $s^3$.
(b): $f_0$ and $f_1$ versus $s$.
\label{FIG-cofu}
}
\end{center}
\end{figure}

The variations of $\alpha$ are essentially linear and given by $\alpha=0.4608+0.44\,c$ excepted at small $c$ where we add the residual correction: $10^{-3} (4.26-9.32\, c)/(1+41.5\,c)$. 
The variations of the other parameters are given in Fig. \ref{FIG-cofu}:
$c$ is essentially proportional to $s^3$, $f_0$ and $f_1$ vary within a factor of two.

\begin{figure}
\begin{center}
\includegraphics[width=0.49\textwidth]{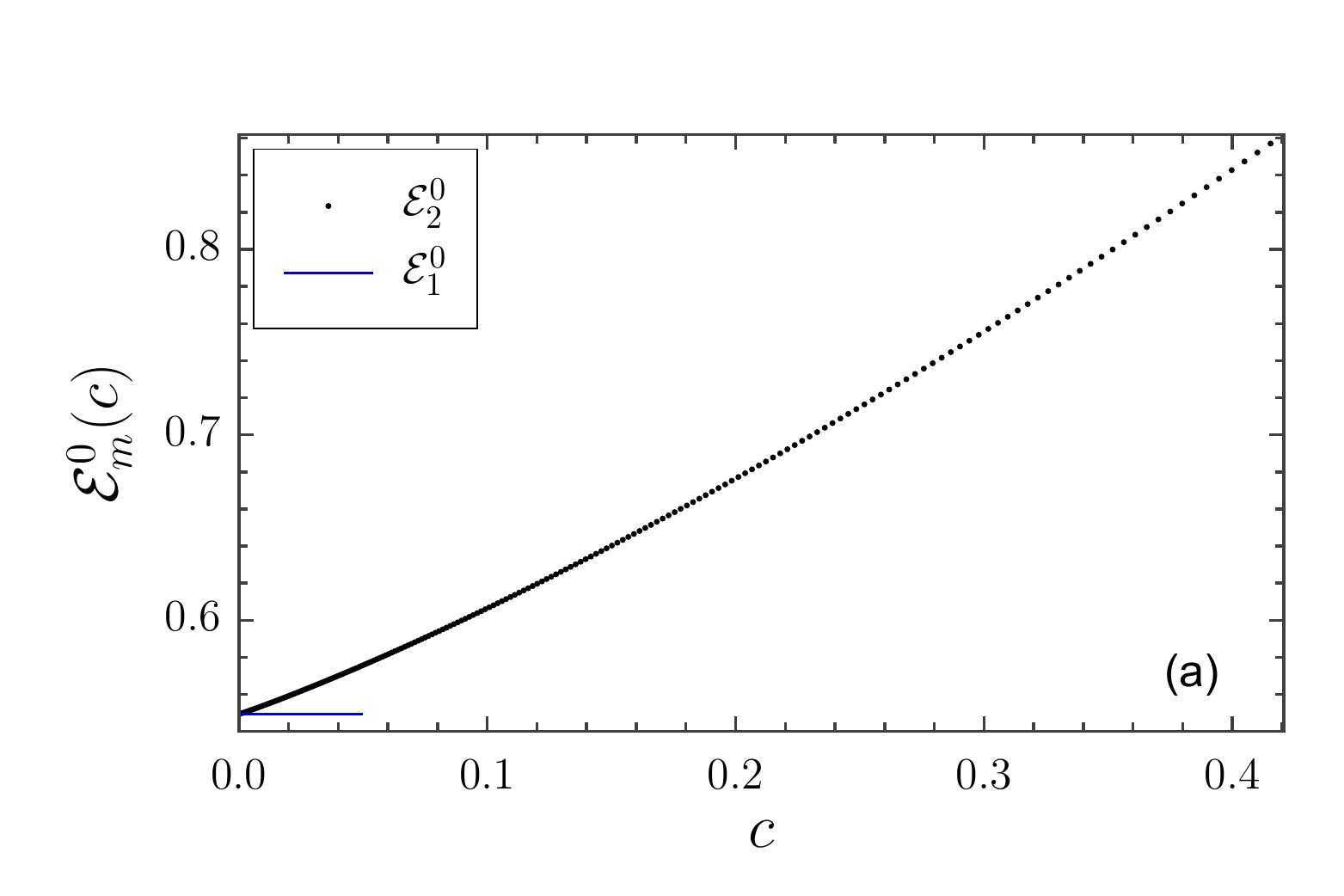}
\includegraphics[width=0.49\textwidth]{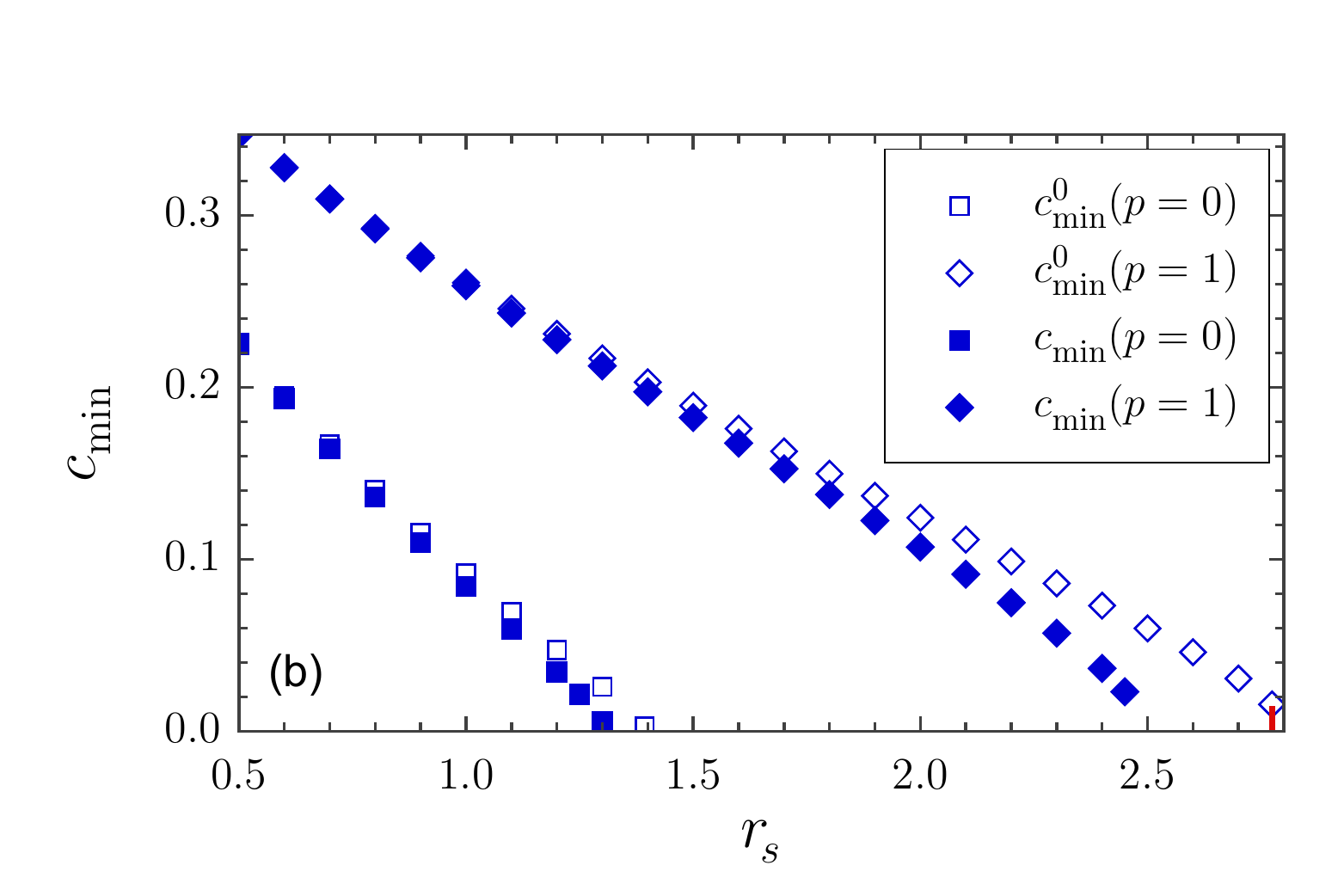}
\caption{
Two-mode model results.
(a):
energy $\E_{2}^{0}(c)$ versus $c\equiv c_1$ the concentration in the first excited state from Eq.(\ref{EQ-defeH0})
compared to the one-mode solution with $\E_1^0 \equiv  \E_{2}^0(c=0)$.
(b): variations of the concentration $\cmin$, versus $r_s$, that minimize $E_2^0$ of Eq.\ref{EQ-e2M1} 
(open symbols) or $E_2$ of Eq.\ref{EQ-e2M2} (full symbol).
Squares (resp. diamond) stand for the unpolarized (resp. polarized) gas.
\label{FIG-TwoModesHartreeEnergy}
}
\end{center}
\end{figure}

Within the Hartree approximation, $\E_{2}^{0}$ in Eq.(\ref{EQ-defeH0}) is still independent of $r_s$ and of the polarization, $p$,
but depends on the concentration $c$. 
(see Fig.\ref{FIG-TwoModesHartreeEnergy}-left).

\subsection{Two-mode exchange term with the Hartree-approximation, $\X_2^0$}
\label{SEC-2MEXCH}

The two-mode exchange term for two modes reads:
\begin{eqnarray}
\label{EQ-defX2}
	\X_{2}(c,r_s)&=&-\frac{r_s^{1/3}}{4\pi}\int_\R\dnu\left[\sum_{a=0}^1 |\tilde\rho_{aa}(\nu)|^2c_a\tilde Y_1\left(\frac{r_s^{1/3} G_p}{\sqrt{c_a}}\nu\right)
			+2|\tilde\rho_{01}(\nu)|^2\tilde Y_2\left(c,r_s^{1/3} G_p\nu\right)\right]
\end{eqnarray}
where $c_0=1-c$ and $c_1=c$.
We refer to Appendix-\ref{APP-Y} for the 
definition and evaluation of the exchange integrals $\tilde Y_1$ and $\tilde Y_2$ which have logarithmic singularities for small $\nu$,
and to Appendix-\ref{APP-exch1M} for the evaluation of the exchange term.

Using the Hartree approximation to determine the shape of the wave functions, the total two-mode energy is approximated by
\begin{eqnarray}
\label{EQ-e2M1}
	E_{2}^0&=&\frac{K_p\left[(1-c)^2+c^2\right]}{r_s^2}+\frac{\E_{2}^{0}(c)}{r_s^{4/3}}+\frac{\X_{2}^{0}(c,\rs)}{\rs}
\end{eqnarray}
At fixed $r_s$, a descent with respect to $c$ allows us to determine
the concentration $\cmina$ which minimizes $E_2^0(c)$.
At small $r_s$, a minimum $\cmina\ne0$ is reached (see Fig.\ref{FIG-TwoModesHartreeEnergy}-right), and
$\cmina$ decreases as $r_s$ increases. The concentration in the excited mode
vanishes at a critical value $r_{s,c}\simeq1.394(1)$ for the unpolarized gas.
For the polarized gas, 
as $r_s$ increases, $c=0$ remains a local minimum.
At $r_s=2.775$, the energy of the two-mode solution with $\cmina\simeq0.015$ crosses the single mode energy.
Thus, within this approximation, we find a first order transition for the polarized gas with a jump in the concentration (see Fig.\ref{FIG-TwoModesHartreeEnergy}-right).

\subsection{Full Minimization}
\label{SEC-2MFULL}
The complete minimization of the total energy with two modes
containing kinetic, Hartree, and exchange energy,  is done by first finding the ground state energy at fixed $\{r_s,c\}$ similar to the single mode case:
\begin{eqnarray}
\label{EQ-e2M2}
	E_{2}&=&\frac{K_p\left[(1-c)^2+c^2\right]}{r_s^2}+\frac{\E_{2}(c,r_s)}{r_s^{4/3}}+\frac{\X_{2}(c,r_s)}{r_s}
\end{eqnarray}
Then, at fixed $r_s$, the minimum, $c_{\rm min}(r_s)$, of the energy is found from a direct Newton-descent on $c$.
The variations of $c_{\rm min}(r_s)$ are close to $c_{\rm min}^0(r_s)$. They only differ significantly close to the transition.
We find a transition at $r_{s,c}=1.30(1)$ for the unpolarized gas and $r_{s,c}=2.50(2)$ for the polarized gas.
In particular, no first order transition subsists for the polarized gas.
The variations of the energy $E_{2}(c_{\rm min})$ versus $r_s$ are
close to $E_{2}^{0}(c_{\rm min}^{0})$ (see Fig.\ref{FIG-energies}). 

\subsection{Existence of three mode solutions}
\label{SEC-3MFULL}
We have further extended the method to study the occupation of  three-modes.
Unfortunately, the series used for the Hartree-approximation do not converge down to $u=0$.
Nevertheless, the solution can be found numerically, and we find the three-mode solution
more stable for $r_s<0.75$ (resp. $r_s<1.6$) for the unpolarized (resp. polarized) gas. 
Since the exchange contribution becomes less important for smaller $r_s$, we 
do not expect significant modifications from  the full HF minimization.

Approaching the high density limit, $r_s \to 0$, we expect an increasing number of
occupied modes.
For $m$ modes, assuming $c_i=1/m$, the kinetic energy is $K_p/(mr_s^2)$, and the dimensionless Hartree energy is a function of $m$ only, $\E_m^0(\{c_i=1/m\}) \equiv F(m)$, as can be seen from Eqs~(\ref{EQ-defEm0},\ref{EQ-defeH0}). Minimizing the 
 total energy, $E_m\approx K_p/(mr_s^2)+F(m)/r_s^{4/3}$,
we can estimate the number of occupied modes in the high density limit: 
\begin{eqnarray}
	m^2F'(m) = K_p r_s^{-2/3}
\end{eqnarray}
Assuming a linear behavior of $F$ for large $m$,
the number of occupied modes diverges as $r_s^{-1/3}$ as $r_s$ approaches zero.

\section{Correlation energy within the local density approximation}
\label{SEC-DFT}

Up to now, we have considered the total energy of the system within the Hartree-Fock approximation
which neglects many-body correlation effects. Within density functional theory (DFT), the
correlation energy per particle for $m$ modes, $\C_m$, defined as the difference between the true total energy 
and the best Hartree-Fock solution, must be a functional of the electronic density only \cite{HohenbergKohn}.
Using the local density approximation (LDA) \cite{KohnSham}, we can write
\begin{equation}
\label{EQ-Cm}
\C_{m} = \int_\R \du \rho(u) \epsilon_{c}^{3D}[r_s^{3D}(u)] 
\end{equation}
where $\epsilon_{c}^{3D}[r_s^{3D}]$ is the correlation energy of the homogenous, three-dimensional electron
gas at the (three-dimensional) density $n^{3D}a_B^3=3 /(4 \pi [r_s^{3D}]^3)$ expressed in terms of the
three-dimensional electron gas parameter $r_s^{3D}$. Using $n^{3D}=\sigma_0 \rho(u)du/dz$ we get
$r_s^{3D}(u)=\left[ 3/ 8\rho(u) \right]^{1/3} r_s^{8/9}$.  
An estimation of the correlation effects is obtained by using the HF density, $\rho(u)$, of the one
and two mode density distribution, together with the Perdew-Zunger \cite{PerdewZunger} parametrization of $\epsilon_{c}^{3D}[r_s^{3D}]$.

\begin{figure}
\begin{center}
\includegraphics[width=0.49\textwidth]{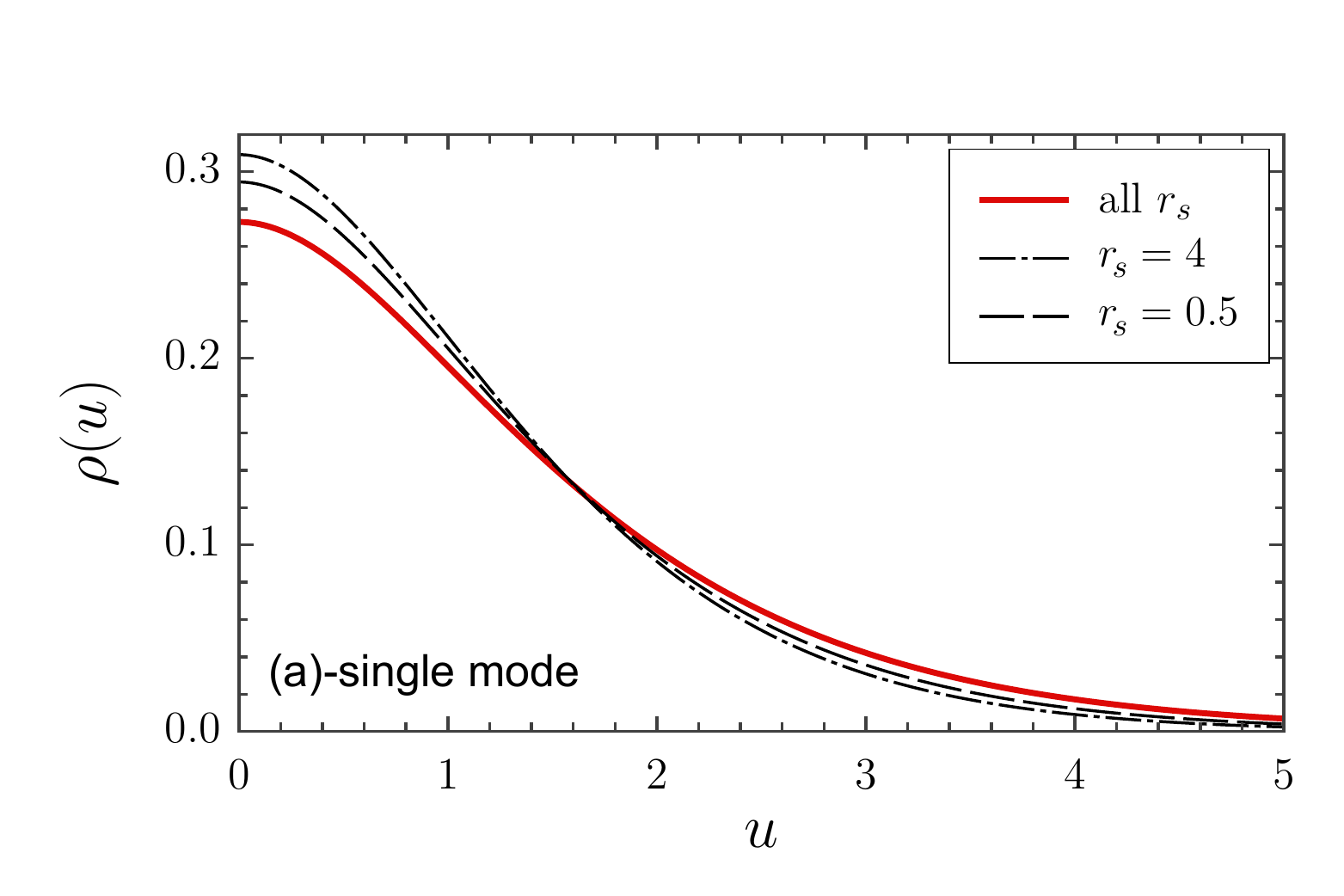}
\includegraphics[width=0.49\textwidth]{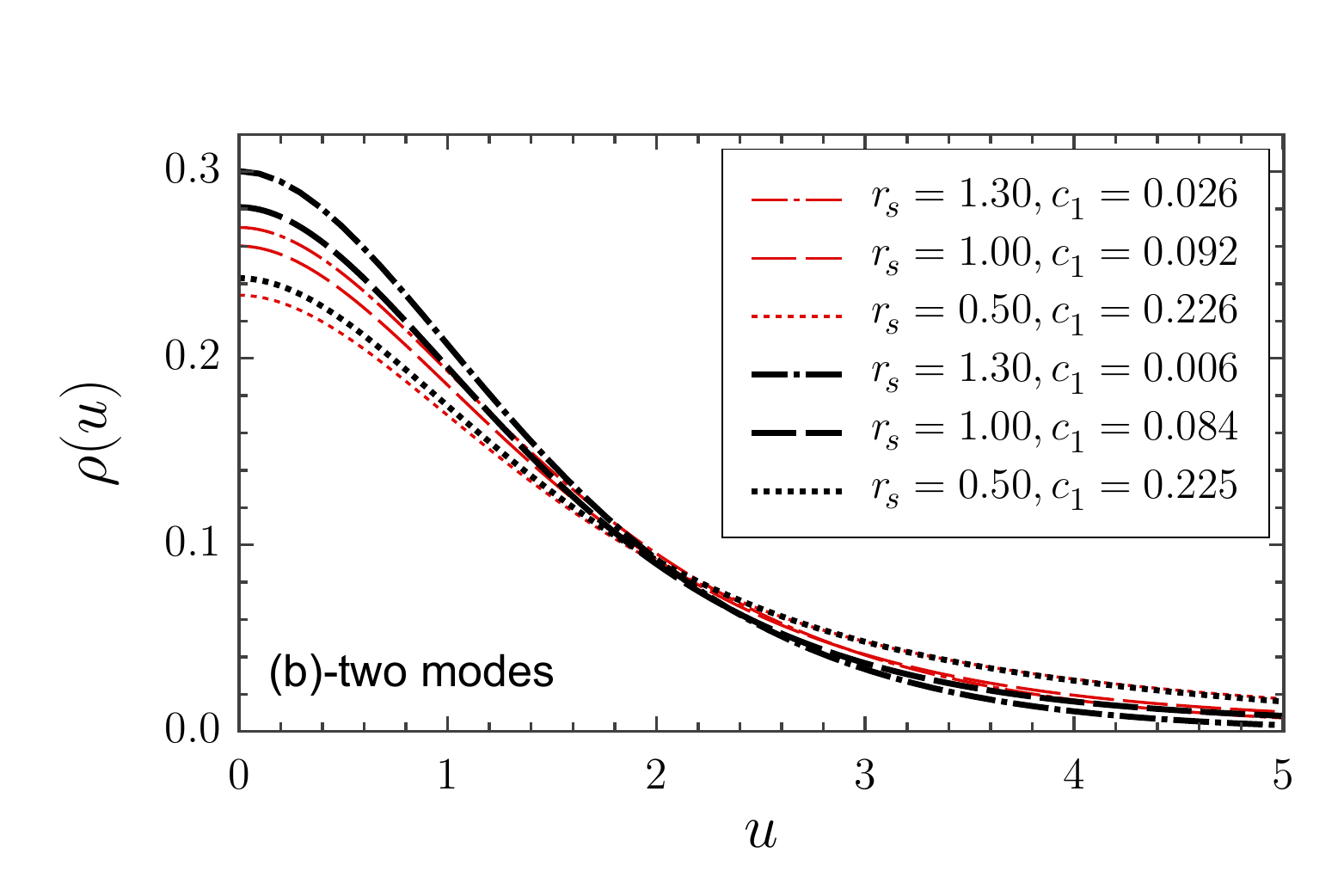}
\caption{
Comparison of the unpolarized charge density profiles $\rho(u=r_s^{1/3}k_Fz)$.
Red lines stand for the analytical Hartree solutions, Eqs. (\ref{EQ-rhos}) and (\ref{EQ-defrho2}) while black lines stand for the optimized densities, as described in \ref{SEC-1MFULL} and \ref{SEC-2MFULL}.
In (b) $c_1$ is the first-excited-mode concentration minimizing the total energy;
at this scale, the corrections coming from LDA-correlation energy are negligible (b).
Notice that the red line at $r_s=1.3$ in (b), with a rather small value of $c_1$  is close to the red line in (a).
\label{FIG-rhocomp}
}
\end{center}
\end{figure}

Around  the transition between one and two excited modes of the unpolarized gas, $r_s \lesssim 1.3$, 
correlations, Eq. (\ref{EQ-Cm}), lower the energy by typically less than $1\%$. Since the
corresponding  total density profiles  (see Fig.~\ref{FIG-rhocomp})  are smoothly varying with $r_s$ and with the concentration in the
first excited state, $c_1$, we do not expect important qualitative and quantitative modifications
 due to correlations in this density region.
Energy minimizations including the LDA-correlation potential, $V_c[\rho(u)]=\delta \C_{m}/\delta \rho(u)$, in the effective Schr\"odinger
equation, confirm that Hartree-Fock accurately describes the high density region where the transition from single to two-mode
occupation of excited modes occurs.

\section{Conclusions}
\label{SEC-Conclusions}

We have studied the model of a quasi-two-dimensional electron gas where electrons are 
confined by a positive
charged background localized in the plane $z=0$. 
Similar to the 2DEG, the electronic density ($\rs$) is the only parameter of the system,
however, 
the phase diagram is different due to possible transition from single to multi-mode occupation in $z$.
Here, we have restricted the discussion to the most simple phases in the metallic regime
neglecting the possibility of 
charge ordering and Wigner crystallization \cite{Wigner,HF}.
Already assuming a simple Fermi liquid wave function
in the high density region, $\rs \to 0$, we have shown that a transition from a single to two
or more occupied modes in the confined direction takes place. Indeed, we expect that close to $\rs=0$ three-dimensional
features to be much more pronounced, as the dominant kinetic energy  favors multi-mode
occupations. 
Further, within HF, the transition between the polarized and unpolarized gas at $r_s\sim4$ occurs
in between the correponding transitions of the 2DEG ($r_s \sim 2$) and the  3DEG ($r_s \sim 5$ )\cite{Needs}.
Similar to 2DEG and 3DEG,
it is likely that the ferromagnetic phase of the Q2DEG is  unstable against Wigner crystallization
within HF, however
correlations are expected to stabilize   
the ferromagnetic fluid phase in higher dimensions \cite{exchange,Drummond}, so that the spin-ordering of the Q2DEG may
essentially differ from that of the 2DEG in the low density region.

Within the Q2DEG, we expect that general aspects of the interplay between correlations and dimensionality
can be studied without the need of a detailed microscopic modeling of a particular experimental device.
This is of particular importance, since  many experimental observations in quasi-two-dimensional electronic
systems reflect strong correlation effects \cite{Spivak}. 
Up to now, precise calculations of correlation effects 
using quantum Monte Carlo methods have
mostly be done for the 2DEG \cite{Tanatar,Drummond, Saverio,mstar}, but
perturbative inclusion of the underlying third dimension have shown to introduce important quantitative changes,
e.g. concerning  the spin susceptibility \cite{Saverio}.  
Within the Q2DEG model non-perturbative calculations are possible, and phases not contained in the 2DEG can be observed.
As a side effect, a quantitative study of the Q2DEG using quantum Monte Carlo methods, may also
provide a reference system, which is strongly inhomogeneous in one direction, so that, within DFT, corrections to the
local density and generalized gradient approximations (GGA)  should be more pronounced, and functionals beyond
LDA/GGA can be tested (see ref.\cite{Kim}). 

\appendix
\section{Properties of the exchange function $\tilde Y$}
\label{APP-Y}
The exchange function is given by the following integral:
\begin{eqnarray}
	\tilde Y(c_a,c_b,\nu)&=&\frac2{\pi^2}
			\int_{|k|^2<c_a} \!\!\!\!\! \dkk
			\int_{|k'|^2<c_b} \!\!\!\!\! \dkkp \frac1{(k-k')^2+\nu^2}
\end{eqnarray}
This function is positive for all $\nu$, even in $\nu$, and satisfies $\tilde Y(c_a,c_b,\nu)=\tilde Y(c_b,c_a,\nu)$
as well as $\alpha\tilde Y\left(\frac {c_a}\alpha,\frac{c_b}\alpha,\frac{\nu}{\sqrt\alpha}\right)=\tilde Y\left(c_a,c_b,\nu\right)$.
We find:
\begin{eqnarray} 
\nonumber
	\tilde Y(c_a,c_b,\nu)
		&=&	\frac4\pi\int_0^{\sqrt{c_a}}\dk k\int_0^{\sqrt{c_b}}\dkp k' 
			\int_0^{2\pi}\!d\theta\,\frac1{k^2+\nu^2+k'^2-2 kk'\cos(\theta)}\\
\nonumber
		&=& 8\int_0^{\sqrt{c_a}}\dk k\int_0^{\sqrt{c_b}}\dkp k' 
			\frac1{\sqrt{(k'^2-k^2+\nu^2)^2+4k^2\nu^2}}\\
\nonumber
		&=& 4\int_0^{\sqrt{c_a}}\dk k\left[
				\arctanh\frac{c_b+\nu^2-k^2}{\sqrt{(k^2-c_b+\nu^2)^2+4c_b\nu^2}}
				-\arctanh\frac{\nu^2-k^2}{k^2+\nu^2}
				\right] \\
\nonumber
		&=& 2\int_0^{c_a}\dk 
			\left[
				\arctanh\frac{c_b+\nu^2-k}{\sqrt{(k-c_b+\nu^2)^2+4c_b\nu^2}}
				-\frac12\ln\frac{\nu^2}{k}
				\right] \\
\label{EQ-Yab}
		&=&X-c_a-c_b-\nu^2+2c_a\ln\frac{X-c_a+c_b+\nu^2}{2\nu^2}
				+2c_b\ln\frac{X+c_a-c_b+\nu^2}{2\nu^2}\qquad\\
\end{eqnarray}
where
\begin{eqnarray} 
\label{EQ-X}
		X&=&\sqrt{(\nu^2+c_a+c_b)^2-4c_ac_b}
\end{eqnarray}
In particular, within the context of the single mode solution it is convenient to introduce the function $\tilde Y_1(\nu)$ given by
\begin{eqnarray}
\nonumber
	\tilde Y_1(\nu/\sqrt c) &=& \tilde Y(c,c,\nu)/c=\tilde Y(1,1,\nu/\sqrt c)\\
\label{EQ-Y1nu}
	\tilde Y_1(\nu)&=&2 t-2-4\ln t\qquad {\rm with }\qquad t^{-1}=\frac12+\frac12 \sqrt{1+\frac{4}{\nu^2}}
	\end{eqnarray}
whereas for the two mode model, we define   $\tilde Y_2(\nu)$
	\begin{eqnarray}
\label{EQ-Y2nu}
\nonumber
	\tilde Y_2(c,\nu) &=&\tilde Y(1-c,c,\nu)\\
	\tilde Y_2(c,\nu)&=&X-1-\nu^2+2(1-c)\ln\frac{X+\nu^2-1+2c}{2\nu^2}
				+2c\ln\frac{X+\nu^2+1-2c}{2\nu^2}
\end{eqnarray}
Both, $\tilde Y_1$ and $\tilde Y_2$, have a logarithmic singularity at $\nu=0$ and behave as $\nu^{-2}$ at large $\nu$:
\begin{eqnarray}
	\tilde Y_1(\nu)&=&-2-4\ln|\nu|+4 |\nu|+{\cal O}(\nu^2)\\
\label{EQ-Y2s}
	\tilde Y_2(c,\nu)&=&-4c\ln|\nu|+2((1-c)\ln(1-c)-(1-2c)\ln(1-2c)-c)+{\cal O}(\nu^2)\\
	\tilde Y_1(\nu)&=&\frac2{\nu^2}-\frac2{\nu^4}+{\cal O}(\nu^{-6})\\
	\tilde Y_2(c,\nu)&=&\frac{2c(1-c)}{\nu^2}-\frac{c(1-c)}{\nu^4}+{\cal O}(\nu^{-6})
\end{eqnarray}

\section{Evaluation of the exchange term}
\label{APP-exch1M}
For the one-mode exchange term of the energy, $\X_1$, we need to evaluate the following integral
\begin{eqnarray}
\label{EQ-eex2}
	\X_1&=&-\frac\beta{2\pi} \int_0^\infty\dnu\tilde\rho(\nu)^2\tilde Y_1(G_p\beta\nu)
\end{eqnarray}
where $\beta=r_s^{1/3}$ and $\tilde\rho(\nu)$ is the Fourier transform of $\rho(u)$.
For $\X_1^0$, the density is defined in Eq. (\ref{EQ-rhos})
and $\tilde\rho(\nu)$ can be computed from
\begin{eqnarray}
	\tilde\rho(\nu)&=&2\int_0^\infty \du \rho(u)\cos(\nu u)
			=2\alpha^4\sum_{k\ge0} (-1)^kf_0^{k+1} \rho_k \frac{2(k+1)\alpha}{4(k+1)^2\alpha^2+\nu^2}
\end{eqnarray}
To remove the logarithmic singularity of the integrand at $\nu=0$, we introduce an auxiliary function $e_1$
\begin{eqnarray}
\label{EQ-eex3}
	\X_1&=&-\frac1{2\pi} \left[e_1(\beta,\tau)+\int_0^\infty\dnu\left[\tilde\rho(\nu)^2 \beta\tilde Y_1(G_p\beta\nu)-\tilde e_1(\nu,\beta,\tau)\right]\right)
\end{eqnarray}
with
\begin{eqnarray}
	\tilde e_1(\nu,\beta,\tau)&=&(e^{-\tau\nu}(1+\tau\nu))^2 \beta(-4\ln(G_p\beta\nu)-2+4G_p\beta\nu)\\
	e_1(\beta,\tau)&=&\int_0^\infty \dnu \tilde e_1(\nu,\beta,\tau)
			=\frac{9G_p\beta^2}{2\tau^2}+\frac \beta\tau\left(-6+5\ln\frac{2\tau}{G_p\beta}+5\gamma\right)
\end{eqnarray}
where $\gamma$ is the Euler constant and $\tau=\sqrt{2\tilde\rho^{(2)}}$ is determined from $\tilde\rho(\nu)=1-\rho^{(2)}\nu^2+{\cal O}(\nu^4)$  ($\rho^{(2)}= 1.617362956587058$ using the solution of Eq. (\ref{EQ-rhos})).
The integral in Eq.~(\ref{EQ-eex3}) is then free of singularities and can be evaluated without major difficulties.

Calculating the first integral in Eq.~(\ref{EQ-defX2}) contributing to the exchange term of two modes, $\X_2$,
we adapt the above procedure for the integrals involving $\tilde\rho_{aa}$ using $\tilde e_2(\nu,\tau)$:
\begin{eqnarray}
	\tilde e_2(\nu,\beta,\tau,c)&=& (e^{-\tau\nu}(1+\tau\nu))^2c\beta(-4\ln(G_p\beta\nu/\sqrt c)-2+4G_p\beta\nu/\sqrt c)\\
	e_2(\beta,\tau,c)&=&c^{3/2}e_1(\beta,\tau\sqrt c)
\end{eqnarray}
The logarithmic singularity in the second contribution containing $\tilde Y_2$ in Eq.(\ref{EQ-defX2}) 
is cancelled by $\tilde\rho_{ab}(\nu)$ which is proportional to $\nu$ at small $\nu$.

Similar auxiliary functions are used to evaluate $V_{aa}^{\rm exc}(u)$, whereas the logarithm singularity of $\tilde Y_2(\nu,c)$ in $V_{ab}^{\rm exc}(u)$ is again cancelled by $\tilde\rho_{ab}(\nu)\propto\nu$.

\section{Details on the numerical minimization scheme}
\label{APP-NUM}
Here we  describe some details on the  numerical minimization of the total Hartree-Fock energy,  Eq.(\ref{EQ-defet}). For simplicity,
we restrict the discussion  to the single mode solution where  the fromal derivative is given by 
$d\psi=\H_d\psi$ ($\H_d=4\H_0{/r_s^{4/3}}+4V_{00}/r_s$).
We proceed using a quadratic minimization scheme.
Let $\psi^{(n)}$ be the solution at step $n$ and  $\{d\psi^{(n-1)},d\psi^{(n)}\}$  the derivatives at step $n-1$ and $n$.
Energies $E(\varepsilon_1,\varepsilon_2)$ are computed at $\psi^{(n)}+\varepsilon_1d\psi^{(n-1)}+\varepsilon_2d\psi^{(n)}$ for the six points $(\varepsilon_1,\varepsilon_2)=(0,0)$, $(\pm\epsilon,0)$, $(0,\pm\epsilon)$ and $(\epsilon,-\epsilon)$.
By assuming a second order polynomial in $\varepsilon_1$ and $\varepsilon_2$, the minimum of  $E(\varepsilon_1,\varepsilon_2)$ is determined analytically, and defines the solution at step $n+1$.

All functions of $u$, e.g. $\psi(u)$, are computed on a grid of $2^p$ points $(i-i_0+1)\delta$ with $i$ from $0$ to $2^p-1$, $i_0=2^{p-1}$ and $\delta=\umax/i_0$. 
Fast Fourier Transform (FFT) are used to compute $\tilde\rho(\nu)$.
In order to achieve good convergence small values of $\delta$ are needed to
accurately calculate  the kinetic energy of the direct (Hartree) potential, whereas a small step in $\nu$ is needed for the exchange energy which implies large values of $\umax$.
We found that $\umax=150$ and $p=10$ are good starting values at sufficiently large value of $c$.
At small $c$, the spatial extension of the excited mode increases significantly which prevents 
accurate solutions for $c \lesssim 10^{-3}$.
Interpolating $\psi(u)$ allows us  to increase $p$ at fixed $\umax$.

\section{Recurrence relation for the two mode Hartree solution}
\label{APP-2Mrec}
We determine the recurrence relation of the series coefficients in the two mode case.
The densities are given by
\begin{eqnarray}
\label{EQ-defrho2}
	\rho=\alpha^4\sum_{k,k'\ge0}\rho_{k,k'}X_0^{2k}X_1^{2k'} 
		&\qquad& \rho_{k,k'}=\rho_{k-1,k'}^{(0)}+s^4\rho_{k,k'-1}^{(1)}\\
\label{EQ-defrhoa}
	\rho_{k,k'}^{(0)}=\sum_{j=0}^{k}\sum_{j'=0}^{k'} a_{j,j'}a_{k-j,k'-j'}
		&& \rho_{k,k'}^{(1)}=\sum_{j=0}^k\sum_{j'=0}^{k'} b_{j,j'}b_{k-j,k'-j'}
\end{eqnarray}
with the convention that $a_{-1,k'}=b_{k,-1}=0$, and
the potential is defined as
\begin{eqnarray}
	v_\rho(u)=-\alpha^2\sum_{k,k'\ge0}v_{k,k'}X_0^{2k}X_1^{2k'}\quad{\rm with}\quad
	v_{k,k'}=\frac{\rho_{k,k'}}{4(k+k's)^2}
\end{eqnarray}
We have
\begin{eqnarray}
	v_\rho\psi_0=-\frac{\alpha^4}{\sqrt{1-c}}\sum_{k,k'\ge0} w_{k,k'}X_0^{2k+1}X_1^{2k'}\quad&{\rm with}&\quad
			w_{k,k'}=\sum_{j=0}^k\sum_{j'=0}^{k'} v_{j,j'}a_{k-j,k'-j'}\\
	v_\rho\psi_1=-\frac{s^2\alpha^4}{\sqrt{c}}\sum_{k,k'\ge0} w_{k,k'}'X_0^{2k}X_1^{2k'+1}\quad&{\rm with}&\quad
			w_{k,k'}'=\sum_{j=0}^k\sum_{j'=0}^{k'} v_{j,j'}b_{k-j,k'-j'}
\end{eqnarray}
Imposing $\psi_0''-(v_\rho+\alpha^2)\psi_0=0$ and $\psi_1''-(v_\rho+s\alpha^2)\psi_1=0$ gives:
\begin{eqnarray}
	\sum_{k,k'\ge0} a_{k,k'}\left[(2k+1+2k's)^2-1\right]X_0^{2k+1}X_1^{2k'}+\sum_{k,k'\ge0} w_{k,k'}X_0^{2k+1}X_1^{2k'}\\
	\sum_{k,k'\ge0} b_{k,k'}\left[(2k+(2k'+1)s)^2-s^2\right]X_0^{2k}X_1^{2k'+1}+\sum_{k,k'\ge0} w_{k,k'}'X_0^{2k}X_1^{2k'+1}
\end{eqnarray}
with the following solution for $(k,k')\ne(0,0)$:
\begin{eqnarray}
	 a_{k,k'}=-\frac{w_{k,k'}}{4(k+k's)(k+1+k's)}\\
	 b_{k,k'}=-\frac{w_{k,k'}'}{4(k+k's)(k+(k'+1)s)}
\end{eqnarray}
Thus, the coefficients $a_{k,k'}$ and $b_{k,k'}$, as well as $\rho_{k,k'}$ and $v_{k,k'}$, are rational functions of $s$ only.


\begin{thebibliography}{99}

\bibitem{Rajagopal} A. K. Rajagopal and J. C. Kimball, Phys. Rev. B {\bf 15}, 2819Ð2825 (1977).

\bibitem{Tanatar} B. Tanatar and D.M. Ceperley, Phys. Rev. B {\bf 39}, 5005 (1989).

\bibitem{Drummond} N. D. Drummond and R. J. Needs, Phys. Rev. Lett. {\bf 102}, 126402 (2009)

\bibitem{AndoFowlerStern} T. Ando, A.B. Fowler, and F. Stern, Rev. Mod. Phys. {\bf 54}, 437 (1982).

\bibitem{Saverio} S. De Palo, M. Botti, S. Moroni, and G. Senatore, Phys. Rev. Lett. {\bf 94}, 226405 (2005).

\bibitem{HohenbergKohn} P. Hohenberg and W. Kohn, {\it
Phys. Rev. } {\bf 136}, B864 (1964).

\bibitem{KohnSham} W. Kohn and L. J. Sham, {\it Phys. Rev.} {\bf 140}, A1133 (1965).

\bibitem{PerdewZunger} J. P. Perdew and A. Zunger, {\it Phys. Rev.} {\bf B 23}, 5048 (1981).

\bibitem{Wigner} E. P. Wigner, Trans. Faraday Soc. {\bf 34}, 678 (1938); Phys. Rev. {\bf 46}, 1002 (1934).

\bibitem{HF} B. Bernu, F. Delyon, M. Duneau, and M. Holzmann, Phys. Rev. {\bf B 78}, 245110 (2008).

\bibitem{Needs} J. R. Trail, M. D. Towler, and R. J. Needs, Phys. Rev. {\bf B 68}, 045107 (2003).

\bibitem{exchange} L. C{\^a}ndido, B. Bernu, and D.M. Ceperley, Phys. Rev. {\bf B 70}, 094413 (2004). 

\bibitem{Spivak} B. Spivak, S. V. Kravchenko, S. A. Kivelson, and X. P. A. Gao, Rev. Mod. Phys. {\bf 82}, 1743 (2010).

\bibitem{mstar} Holzmann, B. Bernu, V. Olevano, R.M. Martin, and D.M. Ceperley, Phys. Rev. {\bf B 79}, 041308(R) (2009).

\bibitem{Kim} Y.H. Kim, I.H. Lee, S. Nagaraja, J.P. Leburton, R.Q. Hood, and R.M. Martin , Phys. Rev. {\bf B 61}, 5202 (2000).

\end{thebibliography}
\end{document}